\documentclass[twocolumn]{aastex61}
\usepackage{amsmath,amstext}
\usepackage[T1]{fontenc}
\usepackage{apjfonts} 
\usepackage[figure,figure*]{hypcap}

\usepackage{graphicx}	
\usepackage{amssymb}	
\usepackage{booktabs}
\usepackage{chemfig}

\def\beq{\begin{equation}}
\def\eeq{\end{equation}}
\def\mnras{MNRAS}
\def\apj{ApJ}

\def\aap{A\&A}
\def\apjl{ApJ Letters}
\def\apjs{ApJS}

\submitjournal{ApJL}
\accepted{2018-01-17}

\shorttitle{Hubble constant from SN Refsdal}
\shortauthors{Vega-Ferrero et al.}

\begin{document}

\title{The Hubble constant  from SN Refsdal}

\correspondingauthor{J. Vega-Ferrero}
\email{vegaf@sas.upenn.edu}

\author[0000-0003-2338-5567]{J. Vega-Ferrero}
\affiliation{Department of Physics and Astronomy, University of Pennsylvania, 209 S. 33rd St, Philadelphia, PA 19104, USA}
\affiliation{IFCA, Instituto de F\'isica de Cantabria (UC-CSIC), Av. de Los Castros s/n, 39005 Santander, Spain}
\author{J. M. Diego}
\affiliation{IFCA, Instituto de F\'isica de Cantabria (UC-CSIC), Av. de Los Castros s/n, 39005 Santander, Spain}
\author{V. Miranda}
\affiliation{Department of Physics and Astronomy, University of Pennsylvania, 209 S. 33rd St, Philadelphia, PA 19104, USA}
\author{G. M. Bernstein}
\affiliation{Department of Physics and Astronomy, University of Pennsylvania, 209 S. 33rd St, Philadelphia, PA 19104, USA}

\begin{abstract}
Hubble Space Telescope observations from December 11 2015 detected the expected fifth counter image of SN Refsdal at $z = 1.49$. In this letter, we compare the time delay predictions from numerous models with the measured value derived by \citet{Kelly2016a} from very early data in the light curve of the SN Refsdal, and find a best value for $H_0 = 64^{+9}_{-11}~\mathrm{km~s^{-1}~Mpc^{-1}}$ (68\% CL), in excellent agreement with predictions from CMB and recent weak lensing data + BAO + BBN (from the DES Collaboration). This is the first constraint on $H_0$ derived from time delays between multiple lensed SN images, and the first with a galaxy cluster lens, so subject to systematic effects different from other time-delay $H_0$ estimates. Additional time delay measurements from new multiply-imaged SNe will allow derivation of competitive constraints on $H_0$.
\end{abstract}


\section{Introduction}
\label{sec:intro}

Galaxy clusters bend the path of photons emitted by distant objects, creating multiple images of the same background source, each with different magnification and arrival times. Time delays between multiple images of the same source depend on the cosmological model, and most notably on the Hubble constant, $H_0$. The potential to constrain $H_0$ with multiple supernova (SN) images was first suggested by~\cite{Refsdal1964}. However, no multiply imaged (and resolved) SN has ever been observed until just recently. In 2014 four counter-images of the same supernova, SN Refsdal \citep{Kelly2015,Rodney2016,Kelly2016a,Kelly2016b}, located at redshift $z=1.49$, were found around a member galaxy in the cluster MACSJ1149.5+2223 \citep[hereafter MACS1149, ][]{Ebeling2001} at redshift $z=0.544$. The predicted time delay between these four images is relatively small (a few days) making them impractical to derive useful constraints on $H_0,$ since the time delays are smaller than the accuracy with which the observed time delay can be determined. Approximately a year after the initial detection of the four supernova images, a fifth counter-image appeared, this one having a considerably longer time delay. The position and the time of reappearance were predicted by different lens models with remarkable precision \citep{Oguri2015,Sharon2015,Diego2016,Treu2016,Jauzac2016}. This accuracy is possible since MACS1149 has been observed with unprecedented detail as part of the \textit{Hubble Frontier Fields} (HFF) program.  MACS1149 contains tens of identified lensed galaxies allowing for detailed modeling of the lens \citep{Jauzac2014,Jauzac2015a,Jauzac2015b,Lam2014,Zitrin2014,Diego2015a,Diego2015b,Diego2016,Kawamata2016,Limousin2016,Mahler2017}.  

The predictions for the SN time delay were based on a set of assumptions, including the value $H_0=70~\mathrm{km~s^{-1}~Mpc^{-1}}$, which was adopted by all teams in their model predictions. Since time delays are inversely proportional to $H_0$, it is possible to constrain the value of $H_0$ directly, as originally suggested by Refsdal. The adopted $H_0$ value lies in between the two of the most precise published estimates. Constraints from the cosmic microwave background (CMB) points towards a {\it low} $H_0 = 66.93 \pm 0.62~\mathrm{km~s^{-1}~Mpc^{-1}}$ \citep{Planck2016} while local measurements favor a {\it high} $H_0 = 73.8 \pm 2.4~\mathrm{km~s^{-1}~Mpc^{-1}}$ \citep{Riess2016} instead. More recently, \citet{Abbott2017} combine intermediate-redshift observations---the Dark Energy Survey Year 1 clustering and weak lensing data with Baryon Acoustic Oscillations (BAO)---with Big Bang Nucleosynthesis (BBN) experiments to obtain $H_0 = 67.2^{+1.2}_{-1.0}~\mathrm{km~s^{-1}~Mpc^{-1}}$ (68\% CL). The difference between intermediate/early and late-time universe cosmologies is of substantial interest. Generalizations of the dark energy phenomenology that predict an equation of state that changes after some transition time could support both results. While \citep{Abbott2017} conclude that the ensemble of $H_0$ data are consistent within uncertainties, additional constraints on the Hubble constant based on measurements at intermediate redshift would provide valuable clarity.

In this letter, we derive an estimate of $H_0$ based on the observed time delay for the SN Refsdal system and an ensemble of lens models derived by different teams that use independent reconstruction methods. Our results provide a separate geometrical inference for $H_0$ \citep{Kundic1997,Sereno2014,Wong2017,Bonvin2017}. Hereafter, we adopt a fiducial cosmological model with $\Omega_m=0.3$ and $\Omega_{\Lambda}=0.7$, which is the cosmology used to infer the lens models. When re-scaling the value of the predicted time delay, the only cosmological parameter being changed is $H_0$ (see section~\ref{sec:hubble} for details). Similar attempts have been made with galaxy-QSO lensing (with $\sim 5$ lensing constraints or less) but never with galaxy clusters where the number of lensing constraints can exceed one hundred. The H0LiCOW-COSMOGRAIL program \citep{Courbin2005,Suyu2017} probably represents the best effort so far to constrain $H_0$ with time delays from multiple lensed QSO. In their most recent work, $H_0$ is constrained with to within $\pm3.8\%$ \citep{Bonvin2017}. Their best inferred value of $H_0$ is in good agreement with the  local  distance  ladder  measurements  of $H_0$. 

The work is organized as follows. Section~\ref{sec:hubble} introduces the time delay formalism and describes the published time delay predictions for SN Refsdal. Section~\ref{sec:bayes} describes the analysis performed to derive the most likely value of $H_0$ from the available data. Finally, sections~\ref{sec:results} and~\ref{sec:conclusions} summarize our results and conclusions.

\section{Hubble constant estimate from SN Refsdal time delays}
\label{sec:hubble}

The time delay $\Delta t$ with respect to an unperturbed null geodesic depends on the angular separation between the image and the source, on the lensing potential at the position of the image, and on the cosmological model through the angular diameter distances. Distances, in turn, depends on the cosmic expansion history of the universe, which is proportional to the Hubble rate,
\begin{equation}
\Delta t(\boldsymbol{\theta}) = \frac{1+z_d}{c}\frac{D_dD_s}{D_{ds}}\left[ \frac{1}{2}(\boldsymbol{\theta} - \boldsymbol{\beta})^2 - \psi(\boldsymbol{\theta}) \right],
\label{eq_timedelay}
\end{equation}
where $\boldsymbol{\beta}$ is the unlensed source position and $\psi(\boldsymbol{\theta})$ is the lens potential at the position of the observed counter-image $\boldsymbol{\theta}$. 
The quantities $D_d$, $D_s$ and $D_{ds}$ are the angular diameter distance to the lens, to the source and between the lens and the source, respectively. These three distances are inversely proportional to $H_0$, and therefore the time delay is also inversely proportional to $H_0$. The factor $D_dD_s/D_{ds}$ encodes the cosmological dependency that, as shown by \citet{Bonvin2017}, is mostly sensitive to $H_0$ and depends weakly on other cosmological parameters. For instance, a change of $10\%$ in the cosmological parameter $\Omega_m$ translates into a change of only $\approx 0.1\%$ in $\Delta t$. Because of this weak dependence on other cosmological parameters, we consider the cosmological model fixed and vary only $H_0$.

The difference in the predicted time delay between two positions in the lens plane depends on a delicate balance between the lensing potential and the relative separations. These terms are typically comparable in magnitude near the minimum of the potential.
Since the term $(\boldsymbol{\theta} - \boldsymbol{\beta})^2$ is straightforward, the uncertainties in the lensing potential are the primary source of systematic errors in the prediction of the time delays, followed by the unknown value of $H_0$. 

Luckily, lensing models for clusters like MACS1149 are constrained by tens of multiply-imaged lensed background galaxies with a wide range of known redshifts that reduce the uncertainties in the lens models \citep{Johnson2016,Acebron2017}.  Traditional time delay measurements based on lensed quasars are based on lens models with typically just one lensed source (the QSO itself) making them more vulnerable to degeneracies like the mass-sheet degeneracy or to projection effects that can involve masses comparable to that of the lens being modeled. Model predictions for time delays are less prone to errors in regions where the number of lensing constraints are more abundant. In the case of MACS1149, the highest density of lensing constraints is found in the vicinity of the multiple supernova images. One should then expect systematics to be relatively small in the case of the SN Refsdal. Finally, the wealth of lens models and predictions for the time delay of the SN Refsdal from multiple investigators allows us to account for the statistical uncertainty in the lens model prediction. We should note however, that some systematic errors may affect all models in a similar fashion (with projection effects being a good example). This will be discussed in more detail later.

\subsection{The case of SN Refsdal}
\label{sc:caseSN}

The SN Refsdal \citep{Kelly2015,Rodney2016,Kelly2016a,Kelly2016b} was the first example of a resolved multiply imaged lensed SN. \citet{Kelly2016a} presents the first estimation of the relative time delay and magnification ratio of S1 (position of knot 1 in the original quadruplet image) and SX (the position at which SN Refsdal reappeared) based on the early light curve of SX. The lensing constraints from the HFF program allowed for a variety of predictions of the time delay and relative magnification of a fifth image. These predictions where made assuming a fiducial cosmological model, needed for computing the distances in Eq.\ref{eq_timedelay}. If the fiducial model assumed the wrong $H_0$ this would translate into a predicted time delay that is biased with respect to the measured one. The predictions ranged from $\approx 8$ months \citep{Sharon2015} to $\approx 1$ year \citep{Diego2016}. Similar predictions extending from $\approx 7.2$ months to $\approx 12.3$ months were later published in \cite{Treu2016} by different teams. SN Refsdal reappeared promptly approximately one year after its first appearance. Overall, the lens models predict reasonably well the time of reappearance of SN Refsdal \citep{Kelly2016a}.

In table~\ref{tb:summary}, we summarize the predicted time delays $\Delta t_{X1}$ between SX and S1, and the magnification ratios, $\mu_{X1} \equiv \mu (X) / \mu (1)$, as derived by the different models \citep{Treu2016}. We also included earlier predictions made by \citet{Diego2016}, labeled as ``Die-16". Both the models ``Die-16" in \cite{Diego2016} and ``Die-a" in \citet{Treu2016} were derived using the same code WSLAP+ \citep{Diego2005,Diego2007} and HFF data, but the model ``Die-16" has higher spatial resolution (and precision) around the positions of the S1-SX counter-images. The model ``Zit-g" in \citet{Treu2016} was found to have numerical errors after the acceptance of that paper, affecting all the ``Zit-g" predictions presented there \citep[see note added in proof in][]{Treu2016}. Therefore, we do not include ``Zit-g" model in the estimation of $H_0$. We have incorporated only models that use data from the HFF program with a large number of constraints. Moreover, we also exclude from the analysis not truly blind predictions, such as the analytic solution for the time delay given in \citet{Jauzac2016} or the updated value of ``Zit-g" model (``Zitrin-c") presented in \citet{Kelly2016a}.

\begin{table}
\begin{center}
\caption{Summary of predicted time delays ($\Delta t_{X1}$) and magnification ratios ($\mu_{X1}$) of the SN image SX relative to image S1 for $H_0=70~\mathrm{km~s^{-1}~Mpc^{-1}}$ presented in \citet{Treu2016}. The uncertainties in both time delays and magnification ratios are only statistical. Note that ``Zit-g" model is not included in the derivation of $H_0$.}
\label{tb:summary}
\begin{tabular}{ccc}
\hline
$\rm{Model}$ & $\Delta t_{X1}$ (days) & $\mu_{X1}$\\
\hline
Die-16 & $376 \pm 25$ & $0.30 \pm  0.05$\\
Die-a & $262 \pm 55$ & $0.31 \pm 0.10$\\
Gri-g & $361^{+19}_{-27}$ & $0.36^{+0.11}_{-0.09}$\\
Ogu-g & $311 \pm 24$ & $0.27 \pm 0.05$\\
Ogu-a & $336 \pm 21$ & $0.27 \pm 0.03$\\
Sha-a & $233^{+46}_{-13}$ & $0.19^{+0.01}_{-0.04}$\\
Sha-g & $277^{+11}_{-21}$ & $0.25^{+0.05}_{-0.02}$\\
\hline
Zit-g & $224 \pm 262$ & $0.31 \pm 0.05$\\
\end{tabular}
\end{center}
\end{table}

The errors listed in table~\ref{tb:summary} are mostly statistical (derived from the dispersion of several models with similar configurations) but in some cases they also include partial systematic errors from varying some of the assumptions made during the lens modeling \citep[see][for a detailed discussion]{Treu2016}. Other sources of systematic effects that would impact the $H_0$ derived from all lens models in the same way were not considered. For example, correlated systematics (i.e, affecting all lens models in a similar fashion) arise from the mass-sheet degeneracy, the effects of line of sight structure, multiplane lensing and unmodeled millilensing. These uncertainties are difficult to quantify since they emerge from assumptions made in the lensing reconstruction and can only be tested with simulations. However, they are generally small in regions on the lens plane where lensing constraints are abundant, as shown by \cite{Meneghetti2016}. Also, \citet{McCully2017} showed how these effects are relatively small in lenses with large Einstein radii and asymmetric image configurations because they are less sensitive to the lens profile degeneracy. SN Refsdal is also in a large spiral galaxy that includes tens of identifiable knots that were used as lensing constraints \citep[see for instance][for a description of these knots]{Diego2016}. \citet{Suyu2013} found that most of the uncertainty in the time delay distance comes from the lens mass model and the line-of-sight contribution. The former diminishes when the number of constraints is large, while the latter can introduce correlations between all model predictions. They quantified the line-of-sight contribution on the time delay uncertainty at the $4.6\%$ level, while other sources of uncertainty, such as the peculiar velocity of the lens, contribute at $<1\%$ level. In this work, we adopt a conservative level of $6\%$ for systematic errors in the time delay predictions \citep[see also][for a similar discussion]{Greene2013}. Recently, \citet{Wilson2017} found that, for the nine time delay lens systems, $H_0$ is overestimated by $11^{+3}_{-2}$\% on average when groups are
ignored.

The lens models assumed a fiducial Hubble constant of $H^{\mathrm{fid}}_0=70~\mathrm{km~s^{-1}~Mpc^{-1}}$, and using the lens geometry data $\mathrm{G}$ each modeler $m$ derived a probability $p_m(\Delta t_{X1}, \mu_{X1} | H^{\mathrm{fid}}_0, \mathrm{G}).$ Since the time delay is inversely proportional to $H_0$ as given by equation~\ref{eq_timedelay}, we can rescale this to any alternative value of $H_0$ via

\begin{equation}
p_m(\Delta t_{X1}, \mu_{X1} | H_0, \mathrm{G}) = 
p_m\left(\frac{H^{\mathrm{fid}}_0}{H_0}\Delta t_{X1}, \mu_{X1} | H^{\mathrm{fid}}_0, \mathrm{G}\right).
\label{eq:tdscaled}
\end{equation}

\section{Bayesian analysis}
\label{sec:bayes}

The time delays $\Delta t_{X1}$ and magnifications $\mu_{X1}$ predicted by the different lens models can be compared with those inferred by \citet{Kelly2016a} from the observed light curve data $LC$ of both SN images. The probability $p_d(\Delta t_{X1},\mu_{X1} | \mathrm{LC})$ derived by \citet{Kelly2016a} shows substantial correlation between $\Delta t_{X1}$ and $\mu_{X1}$ as a consequence of the incompleteness in the light curve data they analyze.  

By re-scaling the predictions as described in equation~\ref{eq:tdscaled}, we can infer the most likely value of $H_0$ that best matches the model predictions with the observations. For this purpose, we adopt a standard Bayesian approach, but keeping in mind that our observational data D are the union $(\mathrm{G},\mathrm{LC})$ of lens geometry and SN light curve data, and both are interpreted in terms of time delay and magnification ratio.  The probability of $H_0$ given the data D is expressed as
\begin{align}
    \label{eq:posterior}
    \mathrm{P} (H_0 | \mathrm{D}) & \propto \mathrm{P}(H_0) ~ \mathrm{P} (\mathrm{D} | H_0) \\
    & = \mathrm{P}(H_0) \int d\Delta t_{X1}\, d\mu_{X1}\, \mathrm{P}(\Delta t_{X1},\mu_{X1} | \mathrm{D}, H_0) \mathrm{P(D)} \\
    & \propto \mathrm{P}(H_0) \int d\Delta t_{X1}\, d\mu_{X1}\,
\nonumber    p_m(\Delta t_{X1},\mu_{X1} | H_0, \mathrm{G}) \\ 
    & \phantom{\propto P(H_0)\qquad} \times p_d(\Delta t_{X1},\mu_{X1} | \mathrm{LC}),
   \label{eq:post_models}
\end{align}
where the prior, $\mathrm{P}(H_0)$, is the credibility of the $H_0$ values without the data D, and the likelihood, $\mathrm{P} (\mathrm{D} | H_0)$, is the probability that the data could be generated by the models with parameter value $H_0$. Expression~\ref{eq:post_models} is basically the product of the observed probability distribution of the observed time delay and magnification times the probability distribution from the individual models. For a particular model, the maximum of the probability is obtained for a value of $H_0$ that maximizes the overlap of the SN light curve data and model probabilities.

For each individual lens model, we assume a bivariate but separable normal distribution for $p_{m,i}(\mu_{X1}, \Delta t_{X1} | H_0)$. The mean values of $\mu_{X1}$ and $\Delta t_{X1}$ for each model are given in table~\ref{tb:summary} along with their statistical uncertainties. In the computation of $p_{m,i}(\mu_{X1}, \Delta t_{X1} | H_0)$, we also take into account that the statistical uncertainties are non-symmetric for three of the listed lens models. For the observational data, we associate a bivariate normal distribution to $p_d(\mu_{X1}, \Delta t_{X1})$ based on the best-fit ellipse to the 68\% CL in Figure 3 of \citet{Kelly2016a} (for brevity, we drop the dependences on $\mathrm{G}$ and $\mathrm{LC}$ from our notation). Both $p_{m,i}(\mu_{X1}, \Delta t_{X1} | H_0)$ and $p_d(\mu_{X1}, \Delta t_{X1})$ are normalized to unity. Note that $p_m(\mu_{X1}, \Delta t_{X1} | H_0)$ depends on $H_0$ as defined in equation~\ref{eq:tdscaled}. On the contrary, the probability distribution of the observational data, $p_d(\mu_{X1}, \Delta t_{X1})$, does not depend on $H_0$. Hereafter, we adopt a very conservative uniform prior $\mathrm{P}(H_0)$ between $H_0=30$ and $100~\textrm{km}\,\textrm{s}^{-1}\,\textrm{Mpc}^{-1}.$

\subsection{Combining models}
\label{sec:posterior}

\begin{figure}
\begin{center}
\includegraphics[width=1.\hsize]{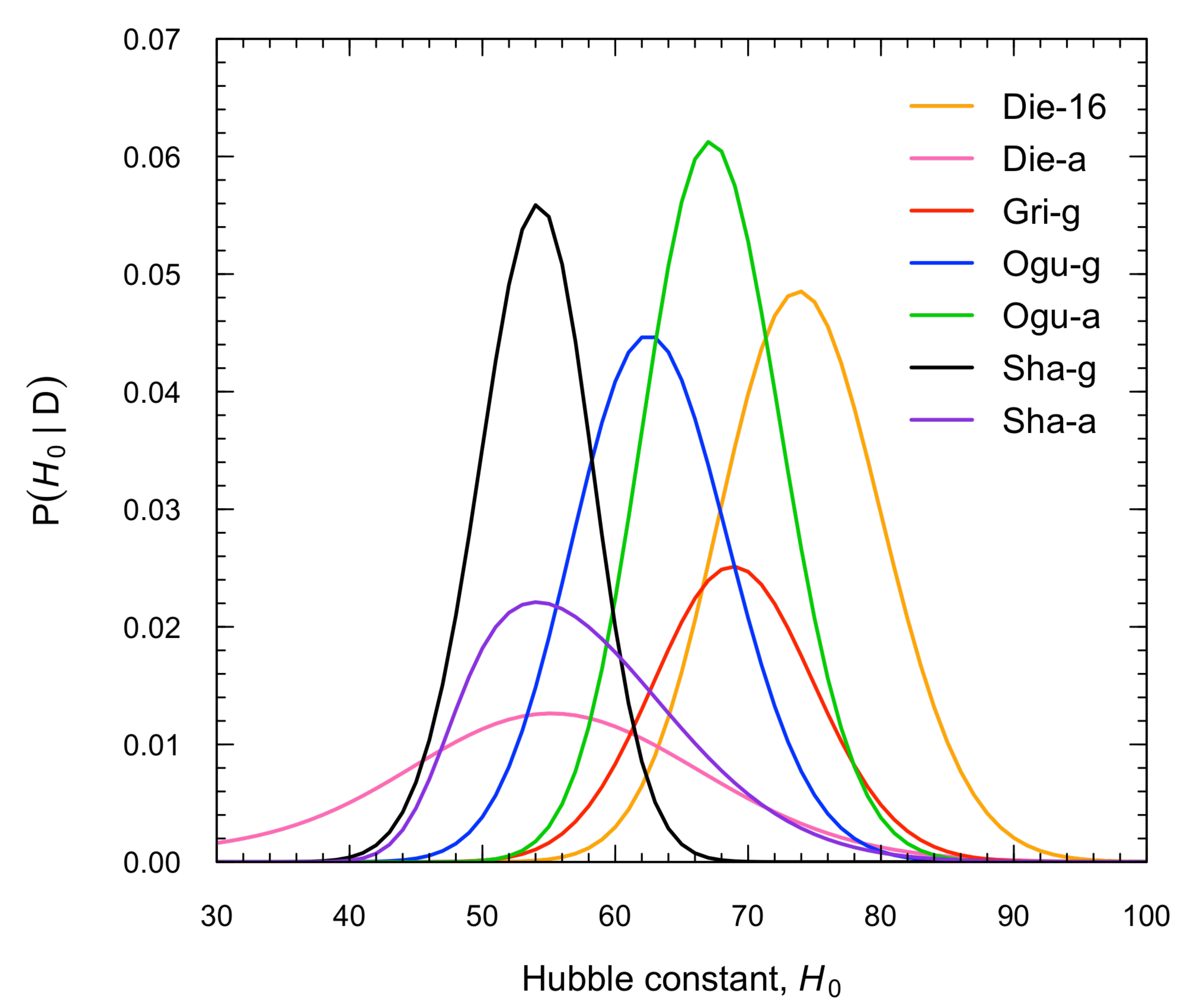}
\caption{Contribution of each lens model prediction (in different colors) to the posterior $\mathrm{P}_+(H_0 | \mathrm{D})$ obtained by equation~\ref{eq:post_models}.}
\label{fig:likelyhood_ind}
\end{center}
\end{figure}

We adopt two strategies for combining the probabilities $p_m$ derived by different lens models, which we can label by $i=1\ldots M$.  A very optimistic view is that each model has errors that are independent and are drawn from an ensemble with zero mean.  In this case we can set
\begin{equation}
p_m(\Delta t_{X1},\mu_{X1}|H_0) \propto \prod_{i=1}^M p_i(\Delta t_{X1},\mu_{X1}|H_0)
\end{equation}
and we will label the resultant posterior derived from equation~\ref{eq:post_models} as $\mathrm{P_{\times}}(H_0 | \mathrm{D}).$

A more conservative (and more realistic) assumption is that only one of the models is correct, with prior probability $q_i$ that model $i$ is the one.  In this case we have
\begin{equation}
p_m(\Delta t_{X1},\mu_{X1}|H_0) = \sum_{i=1}^M q_i p_i(\Delta t_{X1},\mu_{X1}|H_0).
\end{equation}
We will assign equal priors $q_i=1/M$ to each model, such that we effectively average the probabilities of the models, and denote the resultant posterior distribution as $\mathrm{P_+}(H_0 | \mathrm{D}).$  Note that the models do not contribute equally to the posterior: those whose predictions of $\mu_{X1}$ disagree with the measurements of \citet{Kelly2016a} will be downweighted in the integral of equation~\ref{eq:post_models}.

Simplifying assumptions adopted by strong-lensing models about the underlying dark matter can translate into biases in predicting time delays. For example, strong-lensing models may assume symmetries that are not present in the cluster or an absence of small substructures or accurate redshift distribution of the sources.

\begin{figure}
\begin{center}
\includegraphics[width=1.\hsize]{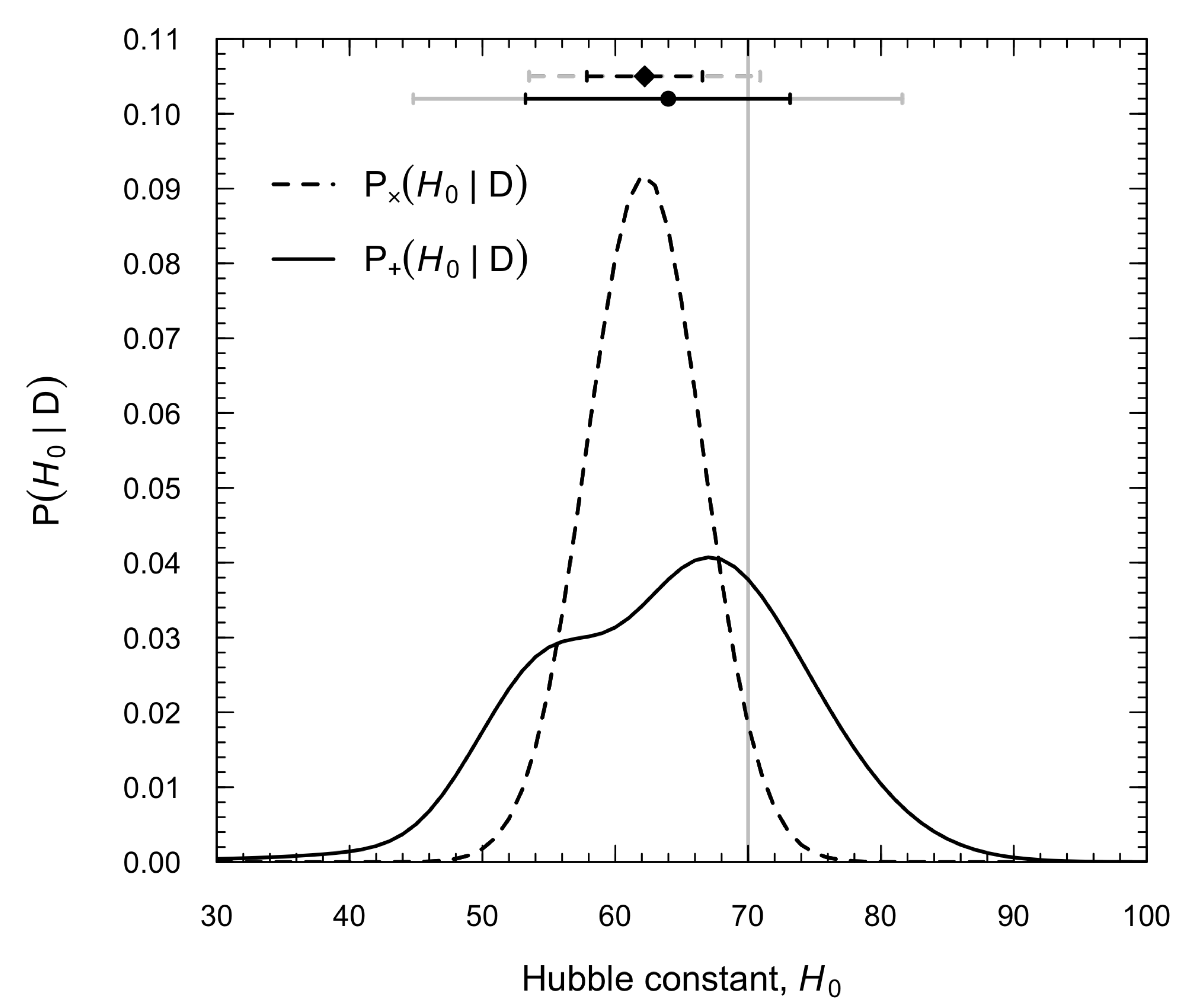}
\caption{Total posterior $\mathrm{P_{\times}}(H_0 | \mathrm{D})$ (dashed line) and $\mathrm{P_{+}}(H_0 | \mathrm{D})$ (solid line). Both curves include a systematic uncertainty at the $6\%$ level added at the end in quadrature to the statistical uncertainty. We explicitly show the median, 68\% CL (black error bars) and 95\% CL (grey error bars) on the top of the figure for both posteriors. The vertical line corresponds to the fiducial $H^{\mathrm{fid}}_0=70~\mathrm{km~s^{-1}~Mpc^{-1}}$ assumed in all the lens models.}
\label{fig:posterior}
\end{center}
\end{figure}

\section{Results}
\label{sec:results}

In figure~\ref{fig:likelyhood_ind} we show the contribution from each model to the total posterior 
$\mathrm{P_{+}} (H_0 | \mathrm{D})$ 
using a flat prior for $H_0$ between $H_0=30$ and $100~\textrm{km}\,\textrm{s}^{-1}\,\textrm{Mpc}^{-1}$.

Figure~\ref{fig:posterior} summarizes our main result for the posterior $\mathrm{P_{\times}}(H_0 | \mathrm{D})$ and $\mathrm{P_{+}}(H_0 | \mathrm{D})$. We have assumed that all model predictions are equal prior validity \citep[but see][for a comparison of the performance of the different lensing reconstruction techniques]{Meneghetti2016}. The median value and 68\% CL for $H_0$ are: $H_0 = 62^{+4}_{-4}~\mathrm{km~s^{-1}~Mpc^{-1}}$ for the $\mathrm{P_{\times}}(H_0 | \mathrm{D})$ posterior; and $H_0 = 64^{+9}_{-11}~\mathrm{km~s^{-1}~Mpc^{-1}}$ for the $\mathrm{P_{+}}(H_0 | \mathrm{D})$ posterior. These values of $H_0$ already include a systematic uncertainty at the $6\%$ level which has been added at the end in quadrature to the statistical uncertainty.

\section{Conclusions}
\label{sec:conclusions}

We constrain, for the first time, the Hubble constant following Refsdal's original idea to use a multiple-lensed SN with measured time delays and precise lens model predictions. By combining the results of multiple lens models, we account for statistical and some systematic errors due to assumptions made during the lens reconstruction. The inferred Hubble constant ranges from $H_0 = 62^{+4}_{-4}~\mathrm{km~s^{-1}~Mpc^{-1}}$ for $\mathrm{P_{\times}}(H_0 | \mathrm{D})$ to $H_0 = 64^{+9}_{-11}~\mathrm{km~s^{-1}~Mpc^{-1}}$ for $\mathrm{P_{+}}(H_0 | \mathrm{D})$. These results are in good agreement with recent constraints from CMB, LSS, and local distance ladders. We use a very weak prior to  better show the sensitivity of the Refsdal data to the parameter $H_0$. Future improved constraints on the observed time delay and magnification will reduce the uncertainty in the Hubble parameter using this technique, and additional estimates derived from different clusters will provide a competitive test for $H_0$.

\acknowledgments

We thank P.L.  Kelly and D. Scolnic for feedback helpful discussions. J.V-F and G.M.B. acknowledge support from the Space Telescope Science Institute (contract number 49726). J.M.D. acknowledges the support of projects AYA2015-64508-P (MINECO/FEDER, UE), AYA2012-39475-C02-01, and the consolider project CSD2010-00064 funded by the Ministerio de Econom\'ia y Competitividad. J.V-F and J.M.D. acknowledge the hospitality of the University of Pennsylvania. V.M. was supported in part by the Charles E.~Kaufman Foundation.


\end{document}